\documentclass[twocolumn,amsmath,amssymb,10pt,superscriptaddress,a4paper,letterpaper,fleqn]{revtex4-1}
\usepackage{amssymb}
\usepackage{epsfig}
\usepackage{graphicx}
\usepackage{dcolumn}
\usepackage{array}
\usepackage{bm}
\usepackage{fancyheadings}
\usepackage{longtable}
\usepackage{multirow}
\usepackage{float}
\usepackage{afterpage}
\usepackage{color}

\setlongtables
\usepackage[breaklinks=true,linkbordercolor={1 1 1}]{hyperref}

\begin{document}
\setcounter{page}{1}

\title{
%% Please do not remove the line below
\qquad \\ \qquad \\ \qquad \\  \qquad \\  \qquad \\ \qquad \\
%% Change title if necessary
Nuclear Level Density within  Extended Superfluid Model with Collective State Enhancement}

\author{V.A. Plujko}
\email[Corresponding author, electronic address:\\ ]{plujko@univ.kiev.ua}
\affiliation{Nuclear Physics Department, Taras Shevchenko National University, Kyiv, Ukraine}
\affiliation{Institute for Nuclear Research, NAS of Ukraine, Kyiv, Ukraine}

\author{O.M. Gorbachenko}
\affiliation{Nuclear Physics Department, Taras Shevchenko National University, Kyiv, Ukraine}

\author{B.M. Bondar}
\affiliation{Nuclear Physics Department, Taras Shevchenko National University, Kyiv, Ukraine}

\author{E.P. Rovenskykh}
\affiliation{Nuclear Physics Department, Taras Shevchenko National University, Kyiv, Ukraine}
\affiliation{Institute for Nuclear Research, NAS of Ukraine, Kyiv, Ukraine}

\date{\today}
%\received{8 March 2013; revised received XX June 2013; accepted XX September 2013}

\begin{abstract}
{
For nuclear level densities, a modification of an enhanced generalized superfluid model with different collective state enhancement factors is studied. An effect of collective states on forming the temperature is taken into account. The ready-to-use tables for the asymptotic value of $a$-parameter of level density as well as for addition shift to excitation energy are prepared using the chi-square fit of the theoretical values of neutron resonance spacing and cumulative number of low-energy levels to experimental values. The systematics of these parameters as a function of mass number and neutron excess are obtained.
The collective state effect on gamma-ray spectra and excitation functions of neutron-induced nuclear reactions is investigated by the use of EMPIRE 3.1 code with modified enhanced generalized superfluid model for nuclear level density.
}
\end{abstract}
\maketitle

%%
%% Headings appear after the \maketitle comand
%%

\lhead{Nuclear level density within  $\dots$}
\chead{NUCLEAR DATA SHEETS}
\rhead{V.A. Plujko \textit{et al.}}
\lfoot{}
\rfoot{}
\renewcommand{\footrulewidth}{0.4pt}

\section{ INTRODUCTION}

The nuclear level density (NLD) $\rho $ is crucial parameter to define characteristics of nuclear decay. The collective states have rather strong effect on NLD, specifically at low excitation energies \cite{73Bjor,83Igna,76Vdov,79Igna,93Igna,09Capo,07Pluj1,07Pluj2}. In fact, enhancement (variation) factor $K_{coll}$ equals to a ratio of NLD with and without ($\rho_{int}$) allowing for collective states.
In the adiabatic approach, enhancement factor $K_{coll}$ is a product of vibrational $K_{vibr}$ and
rotational $K_{rot}$ enhancement factors that take into account the change of level densities due to presence of vibrational
and rotational  states respectively.
Up to now, there exist problems in estimation of these variation factors. Specifically, there is rather big uncertainties in estimation of the magnitude of the $K_{vibr}$  and different approaches \cite{73Bjor},\cite{79Igna,93Igna,09Capo,07Pluj1,07Pluj2} lead to different values of the $K_{vibr}$. Variations of the $K_{vibr}$ with excitation energy are strongly dependent on vibrational state damping width, and this is unresolved task too. Microscopic description of the vibrational state relaxation still remain to be answered and different phenomenological approaches of allowing for vibrational state damping are used.
In this contribution, the phenomenological methods of description of vibrational state contribution into NLD are tested with  implementation of the Enhanced Generalized Superfluid Model (Empire Global Specific Model, EGSM)\cite{07Herm}  for description of level densities  of intrinsic and rotational states.

\section{THE METHODS OF VIBRATIONAL ENHANCEMENT FACTOR CALCULATIONS}

The simple methods for calculations  of the vibrational enhancement factor $K_{vibr}=\rho /\rho _{int} $ are based on the  saddle-point method with the partition function $Z=Z_{int} \cdot \Delta Z$, where $\Delta Z$ is co-factor resulted from collective coherent interaction forming vibrational states ($\Delta Z$ is named below as vibrational co-factor of partition function). Generally, collective states change the temperature of intrinsic states $T_{int}$.
As it is shown in \cite{07Pluj1,07Pluj2}, the nuclear temperature $T$ is equal to the temperature of intrinsic states $T_{int}$
in the first order on the variation $\delta T=T-T_{int}$.

The simple phenomenological methods for calculations of the vibrational enhancement factor were proposed in \cite{73Bjor},\cite{93Igna,09Capo,07Pluj1,07Pluj2,05Pluj1}. They are based on different phenomenological extensions of the boson expression ($Z_{bos}$) for the vibrational co-factor $\Delta Z$\cite{07Pluj1,07Pluj2} with an approximation $K_{vibr}=\Delta Z(T_{int})$.
Among these methods, there is damped occupation number approach ($K_{vibr}=K_{DN}) $\cite{79Igna,93Igna,09Capo},  liquid drop prescription with  the temperature damping ($K_{EM}$) \cite{07Herm} and simplified version \cite{07Pluj1,07Pluj2}  of the response function method  \cite{05Pluj1}. According to this last approach, the $\Delta Z$ is taken as a ratio of boson partition functions with averaged occupation numbers (BAN approach): $\Delta Z=\prod _{L}((1+<n(\omega _{L} )>)/$ $(1+<n(\tilde{\omega}_{L} )>))^{2L+1}  \equiv \Delta Z_{BAN}$. The quantities $<n(\omega_L )>$ are boson occupation numbers averaged over the collective motion period  \cite{07Pluj1,07Pluj2}: $<n(\omega _{L} )>=$ ${\hbar \omega _L }{\left(1-\exp (-2\pi\Gamma _{L}/\hbar\omega_L) \right)}/({2\pi \Gamma _L }{(\exp (\hbar\omega_L/T)-1)})$ with $\Gamma _{L} $ for damping width of vibrational state of multipolarity $L$ with characteristic frequency $\omega _{L} =E_{L} /\hbar $ and $E_{L} $ is the energy of this vibrational state; $\tilde\omega_L$ is a frequency of corresponding $1p1h$ state. The damping width is determined by
collective relaxation time $\tau _{C}$ that results from retardation effects during two-body collisions:
$\Gamma _{L}(\omega) =2\hbar /\tau _{C} (\omega  ,T)$ \cite{07Pluj1,07Pluj2}.

\section{RESULTS OF CALCULATIONS AND CONCLUSIONS}

We use the following expression of NLD for states with excitation energy $U$  and spin $J$:
$ \rho (U,J)=\bar\rho _{EM}(U,J)\cdot K_{vibr}$,
where $\bar\rho_{EM}$ is level density that takes into account excitations of  intrinsic and rotational states. It is calculated using the EGSM of the EMPIRE 3.1 \cite{07Herm}.
The $\bar\rho_{EM}$ is a function of the asymptotic value $\tilde a$ of the $a$-parameter of level density
and $\tilde\delta_{shift}$ is an additional shift of excitation energy.

 For NLD, the expression $\rho=\bar\rho_{EM}K_{vibr}$ is used with different $K_{vibr}$
and quantities  $\tilde a$, $\tilde\delta_{shift}$ are considered as the parameters. The  $\tilde a$ is obtained from fitting of average theoretical NLD $<\rho>=1/D$ to the $1/D_{exp}$  corresponding experimental data on $s$-resonance spacing \cite{09Capo}. Shift parameters are obtained from fit of the experimental values of cumulative numbers  $N_{cum}^{exp}$ of low-lying discrete levels to theoretical values $N_{cum}=\int _{0}^{U_{cum} }\sum_{J}\rho (U,J)dU$ with previously determined $\tilde a$.
This approach is referred below as modified EGSM.

The ready-to-use table of the  parameters $\tilde{a}$ and  $\tilde\delta _{shift}$ was prepared with the use of this approach for 291 nuclei and BAN approach for $K_{vibr}$. The following systematics were also obtained for the parameters $\tilde{a}$, $\tilde\delta _{shift}$ with dependence on mass number $A$ and neutron excess $I=(N-Z)/A$:
$\tilde{a} =\alpha _{V} A {\rm (1+}\alpha _{VI} I^{{\rm 2}} {\rm )+}\alpha _{S} A^{2/3}({\rm 1+}\alpha _{SI} I^{{\rm 2}} {\rm )+}\alpha _{C} Z^2/A^{1/3}$, (MeV$^{-1}$),
$\tilde\delta _{shift} =\delta _{1} {\rm (1+}\delta _{1I} I^{{\rm 2}} {\rm )+}\delta _{2} A({\rm 1+}\delta _{2I} I^{{\rm 2}} {\rm )+}\delta _{3} E_{2_{1}^{+} }$, (MeV),
where $E_{2_{1}^{+} } $- energy of the first $2^+$  collective state in MeV.
The values of the parameters of systematics and their uncertainties are presented in Table \ref{Table1}.

\begin{table}[!htb]
\caption{The coefficients of systematics for $\tilde a$ and $\tilde\delta _{shift}$}
\label{Table1}
\begin{tabular}{c|c|c|c|c|c} \hline \hline
$\tilde{a}$ & $\alpha _{V} $ & $\alpha _{VI} $ & $\alpha _{S} $ &  $\alpha _{SI} $ & $\alpha _{C} $ \\ \hline
 & 0.527\textit{6} & 4.50\textit{6} & -1.17\textit{2} &  13.3\textit{2} & -0.0447\textit{6} \\ \hline
$\tilde\delta _{shift} $ & $\delta _{1} $ & $\delta _{1I} $ & $\delta _{2} $ & $\delta _{2I} $ & $\delta _{3} $ \\ \hline
 & 1.37\textit{2} & -11.7\textit{5} & 0.00000003\textit{1}  & 76503\textit{23974} & -0.697\textit{9} \\ \hline \hline
\end{tabular}
\end{table}

\begin{table}[!htb]
\caption{The comparison of the ratios $\sum _{i=1}^{N_{nucl} }\chi _{i}^{2} (K_{vibr} ) /\sum _{i=1}^{N_{nucl} }\chi _{i}^{2} (K_{vibr} =1)$. }
\label{Table2}
\begin{tabular}{c|c|c|c|c} \hline  \hline
Data & $K_{EM}$ & $K_{DN}$ & $K_{BAN}$ & $K_{BANT}$ \\ \hline
\cite{04Agva}& 1 & 0.9  & 1.0 & 0.9 \\ \hline
\cite{07Sukh} & 4.0 & 1.7 & 1.5 & 1.6 \\ \hline
\cite{09Zhur}  & 1.5 & 5.5  & 0.9 & 0.6 \\ \hline
average & 2.2 & 2.7  & 1.1 & 1.0 \\ \hline  \hline
\end{tabular}
\end{table}

In Table \ref{Table2} are given the ratios $\sum _{i=1}^{N_{nucl} }\chi _{i}^{2} (K_{vibr} ) /\sum _{i=1}^{N_{nucl} }\chi _{i}^{2} (K_{vibr} =1)$
of chi-square deviations $\chi _{i}^{2}(K_{vibr}) =\sum _{j=1}^{n_{i} }(\rho _{theor,i} (U_{j} )-\rho _{\exp ,i} (U_{j} ))^{2} /n_{i}$ of theoretical NLD  within modified EGSM with different $K_{vibr}$ from experimental data of teams from Oslo \cite{04Agva}, Dubna \cite{07Sukh} and Obninsk \cite{09Zhur}. The $n_i$ is number of experimental data for nucleus $i$ and $N_{nucl}$ is number of nuclei.
For approximation BANT,  vibrational co-factor $\Delta Z _{BAN}$ of partition function was used and variation of the temperature $\delta T=T-T_{int}$ was taken into account.
The partition function of back-shifted Fermi gas model (BSFG) was used for intrinsic states.
For used experimental data, relative chi-square deviations for BAN and BANT approaches are less than for  other models with vibrational enhancement. The BAN approach is appeared to be more preferable in comparison with BANT
because calculations within BANT are rather complicated.

\begin{figure}[!htb]
\includegraphics[width=0.9\columnwidth]{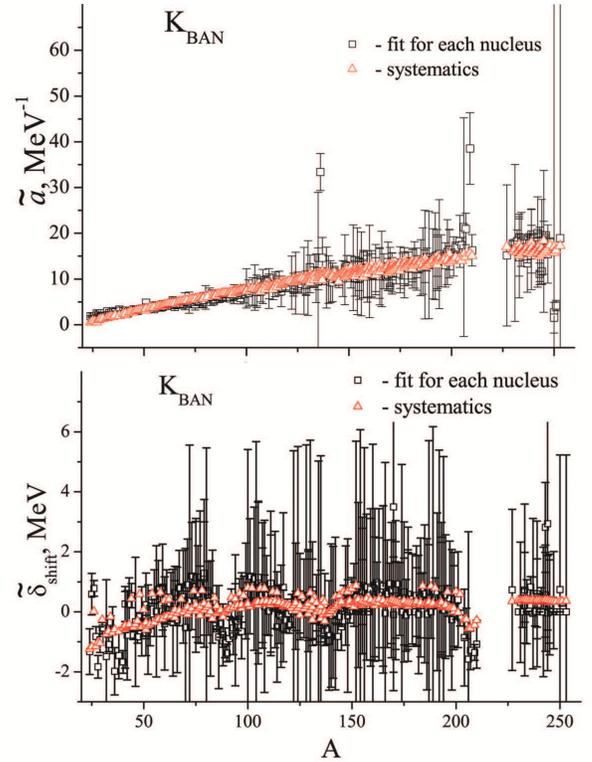}
\caption{The comparison of $\tilde{a}$  and  $\tilde\delta_{shift} $  for each nucleus with the systematics.}
\label{fig1}
\end{figure}

\begin{figure}[!htb]
\includegraphics[width=0.9\columnwidth]{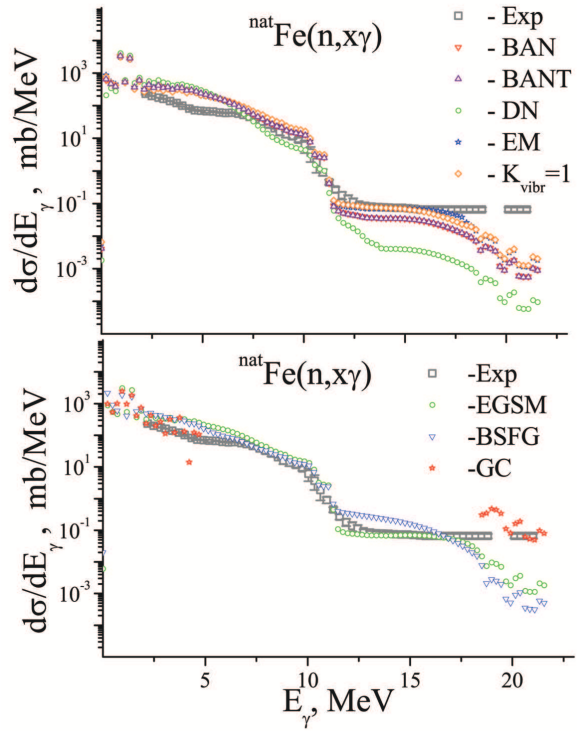}
\caption{Dependence of gamma-ray spectra on gamma-ray energy  for $^{nat} {\rm Fe}(n,x\gamma )$ reaction, $E_{n} =14.1\, \, MeV$. }
\label{fig2}
\end{figure}

The comparisons of  $\tilde{a}$ and  $\tilde\delta _{shift} $  for each nucleus with their systematics are shown on Fig.\ref{fig1}.

Figure \ref{fig2} presents gamma-ray spectra for $(n,x\gamma )$ reactions at $E_{n} =14.1\, \, MeV$ on $ {nat} {\rm Fe}$.
Theoretical spectra were calculated  using modified EGSM as well as Gilbert-Cameron (GC)  model and back-shifted Fermi gas model. Experimental data were taken from \cite{11Bond}. It can be seen, that scatter of gamma-ray spectra calculated within
modified EGSM  are the same order as scatter of the  spectra calculated using other NLD models.

\begin{figure}[!htb]
\includegraphics[width=0.9\columnwidth]{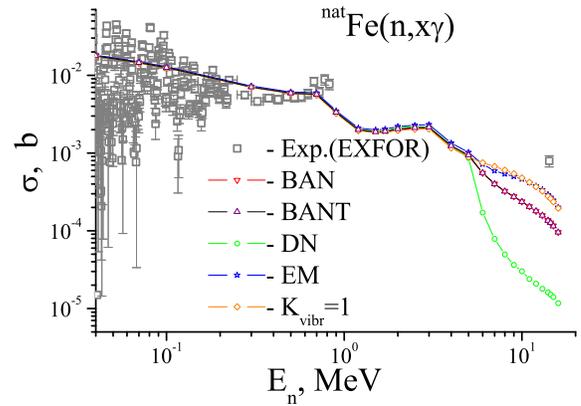}
\caption{Excitation function of reaction $(n,x\gamma)$ on $^{nat}Fe$.}
\label{fig3}
\end{figure}

Figure \ref{fig3} shows the excitation functions of reaction $(n,x\gamma)$ on $^{nat}Fe$.
Experimental data are taken from EXFOR data library.
Theoretical excitation functions are calculated using modified EGSM with different $K_{vibr}$.
One can see that shape and values of excitation function and gamma-ray spectra are sensitive to choice
of vibrational enhancement factor.

For modified EGSM, approximation of boson partition function with average occupation numbers (BAN) can
be considered as the most appropriate approach for calculation of the vibrational enhancement factor.
The comparison between theoretical calculations and experimental data also shows that within BAN
approach $K_{vibr}(S_n) \sim 2 \div 5$ for
$A \sim 100$. These values of enhancement factor are in agreement with results of  microscopic quasiparticle-phonon
model\cite{76Vdov}.


\begin{thebibliography}{9}

\bibitem{73Bjor}
S. Bjornholm {\it et al.}, {\sc Proc. of Symposium Rochester, New-York, 13-17 August., 1973} 367 (1973).

\bibitem{83Igna}
A.V. Ignatyuk, {\sc Statistical Properties of Excited Atomic Nuclei, IAEA, INDC-233(L),} (1985).


\bibitem{76Vdov}
A.I. Vdovin {\it et al.}, {\sc Phys. Elem. Particles Atom. Nuclei} {\bf 7},  952 (1976).

\bibitem{79Igna}
A.V. Ignatyuk {\it et al.}, {\sc  Sov.J.Nucl.Phys.} {\bf 29}, 450 (1979).

\bibitem{93Igna}
A.V. Ignatyuk {\it et al.}, {\sc Phys. Rev. C} {\bf 47}, 1504 (1993).

\bibitem{09Capo}
R. Capote {\it et al.}, {\sc Nucl. Data Sheets} {\bf 110}, 3107 (2009);  http://www-nds.iaea.org/RIPL-3/.


\bibitem{07Pluj1}
V.A. Plujko {\it et al.}, {\sc Phys.Atom.Nucl.}  {\bf 70}, 1643 (2007).

\bibitem{07Pluj2}
V.A. Plujko {\it et al.}, {\sc  Int.Jour.Mod. Phys. E}  {\bf 16}, 570 (2007).

\bibitem{05Pluj1}
V.A. Plujko {\it et al.}, {\sc AIP Conf. Proc.} {\bf 769},  1124 (2005).

\bibitem{07Herm}
M. Herman {\it et al.}, {\sc  Nucl. Data Sheets} {\bf 108}, 2655 (2007).; http://www.nndc.bnl.gov/empire/ .

\bibitem{04Agva}
U. Agvaanluvsan {\it et al.}, {\sc Phys. Rev.C} {\bf 70}, 054611 (2004); http://ocl.uio.no/compilation/ .

\bibitem{07Sukh}
A.M. Sukhovoj {\it et al.}, {\sc  Proc. Int. Conf. ISINN-15}, Dubna, May 2007. 92 (2007)

\bibitem{09Zhur}
B.V. Zhuravlev, {\sc IAEA, INDC(NDS)-0554, Distr. G+NM} (2009)

\bibitem{11Bond}
V.M. Bondar {\it et al.}, {\sc Proc. Int. Conf. ISINN-18}, Dubna, May 26-29, 2010.  135 (2011)

\end{thebibliography}
\end{document}